\documentclass[reprint,prl,aps,superscriptaddress]{revtex4-2}
\usepackage{times}
\usepackage{graphicx}
\usepackage{amsmath, braket, amsfonts}
\usepackage{amssymb}
\usepackage{sidecap}
\usepackage[normalem]{ulem}
\usepackage{bm, color, ulem}
\usepackage{xcolor}
\usepackage{ragged2e}
\usepackage{wrapfig,lipsum,booktabs}
\usepackage{lineno}

\usepackage{siunitx}

\newcommand{\be}{\begin{equation}}
\newcommand{\ee}{\end{equation}}

\DeclareUnicodeCharacter{03BD}{{$\nu$}}

\makeatletter
\def\maketitle{
\@author@finish
\title@column\titleblock@produce
\suppressfloats[t]}
\makeatother

\begin{document}
\title{Nanoscale infrared and microwave imaging of stacking faults in multilayer graphene}
\author{Ludwig Holleis}  
\affiliation{Department of Physics, University of California at Santa Barbara, Santa Barbara CA 93106, USA}
\author{Liam Cohen}  
\affiliation{Department of Physics, University of California at Santa Barbara, Santa Barbara CA 93106, USA} 
\author{Noah Samuelson}  
\affiliation{Department of Physics, University of California at Santa Barbara, Santa Barbara CA 93106, USA} 
\author{Caitlin L. Patterson}  
\affiliation{Department of Physics, University of California at Santa Barbara, Santa Barbara CA 93106, USA} 
\author{Ysun Choi}  
\affiliation{Department of Physics, University of California at Santa Barbara, Santa Barbara CA 93106, USA} 
\author{Marco Valentini}  
\affiliation{Department of Physics, University of California at Santa Barbara, Santa Barbara CA 93106, USA} 
\author{Owen Sheekey}  
\affiliation{Department of Physics, University of California at Santa Barbara, Santa Barbara CA 93106, USA} 
\author{Youngjoon Choi} 
\affiliation{Department of Physics, University of California at Santa Barbara, Santa Barbara CA 93106, USA} 
\author{Jiaxi Zhou}  
\affiliation{Department of Physics, University of California at Santa Barbara, Santa Barbara CA 93106, USA} 
\author{Hari Stoyanov}  
\affiliation{Department of Physics, University of California at Santa Barbara, Santa Barbara CA 93106, USA} 
\author{Takashi Taniguchi}
 \affiliation{International Center for Materials Nanoarchitectonics,
 National Institute for Materials Science,  1-1 Namiki, Tsukuba 305-0044, Japan}
 \author{Kenji Watanabe}
 \affiliation{Research Center for Functional Materials,
 National Institute for Materials Science, 1-1 Namiki, Tsukuba 305-0044, Japan}
\author{Qichi Hu}  
\affiliation{Bruker Nano Surfaces, Santa Barbara CA 93117, USA.}
\author{Jin Hee Kim }  
\affiliation{Bruker Nano Surfaces, Santa Barbara CA 93117, USA.}
\author{Cassandra Phillips}  
\affiliation{Bruker Nano Surfaces, Santa Barbara CA 93117, USA.}
\author{Peter De Wolf}  
\affiliation{Bruker Nano Surfaces, Santa Barbara CA 93117, USA.}
\author{Andrea F. Young}
\email{andrea@physics.ucsb.edu}
 \affiliation{Department of Physics, University of California at Santa Barbara, Santa Barbara CA 93106, USA}
\date{\today}

\begin{abstract}
    Graphite occurs in a range of metastable stacking orders characterized by both the number and direction of shifts between adjacent layers by the length of a single carbon-carbon bond. At the extremes are Bernal (or ``ABAB...'') stacking, where the direction of the interlayer shift alternates with each layer, and rhombohedral (or ``ABCABC...'')  stacking order where the shifts are always in the same direction.  However, for an N-layer system, there are in principle $N-1$ unique metastable stacking orders of this type.  Recently, it has become clear that stacking order has a strong effect on the low energy electronic band structure with single-layer shifts completely altering the electronic properties.
    Most experimental work has focused on the extremal stacking orders in large part due to the difficulty of isolating and identifying intermediate orders. Motivated by this challenge, here we describe two atomic force microscopy (AFM) based techniques to unambiguously distinguish stacking orders and defects in graphite flakes. 
    Photo-thermal infrared atomic force microscope (AFM-IR) is able to distinguish stacking orders across multiple IR wavelengths and readily provides absolute contrast via IR spectral analysis. 
    Scanning microwave impedance microscopy (sMIM) can distinguish the relative contrast between Bernal, intermediate and rhombohedral domains. 
    We show that both techniques are well suited to characterizing graphite van der Waals devices, providing high contrast determination of stacking order, subsurface imaging of graphene flakes buried under a hexagonal boron nitride (hBN) dielectric layer, and identifying nanoscale domain walls.  Our results pave the way for the reliable fabrication of graphene multilayer devices of definite interlayer registry.
\end{abstract}

\maketitle

\section{Introduction}
The band structure of few-layer graphene depends strongly on its stacking order. For example,
Rhombohedral graphite stacking has a single low energy band with an approximate k$^N$ band dispersion, with N the number of graphene layers.
Recently, rhombohedral N-layer graphene (RNG) has attracted significant attention within the family of graphene systems. 
Within its high density of states regime close to the band edge, broken isospin symmetries\cite{Shi2020,Zhou2021a, Zhou2022, Han2023,Holleis2025}, superconductivity\cite{Zhou2021, Zhou2022, Zhang2023,Holleis2023,Li2024,Patterson2024,Choi2024,Yang2024,Han2024a}, intervalley coherent states\cite{Arp2023,Liu2024} as well as integer and fractional Chern insulators\cite{Han2024,Lu2024,Choi2024,Lu2024a,Xie2024,Aronson2024,Han2024b} have all been observed. In contrast, the electronic band structure of Bernal graphite, the energetically favorable stacking order, typically features multiple bands at low energy, often including strongly dispersing bands, and similar effects have not, to date, been observed experimentally. 
This motivates the development of high throughput techniques that may aid in the fabrication of devices consisting of pure, metastable rhombohedral structure, as well as the identification of intermediate stacking orders that may host exotic correlated electronic states of matter.

Previously, RNG has been identified by angle resolved photoemission\cite{Ohta2007}, infrared \cite{Mak2010} and Raman\cite{Lui2011a,Cong2011} spectroscopy.
IR spectroscopy has proven to be a particularly powerful tool which harnesses differences in the electronic bands below 0.8 eV to generate optical contrast between different stacking orders\cite{Mak2010}.
Recent efforts have used this effect to directly image graphite stacking faults using a high sensitivity InGaAs IR camera\cite{Lu2024a,Feng2024}.
However, the large area IR-images provide only a relative contrast between stacking orders; while this is effective for identifying purely rhombohedral domains in the few layer limit, it fails in terms of both absolute certainty of identification for $N>3$ layers. 
In addition, near-field imaging techniques have shown that stacking faults, such as domain walls, can occur with lateral size far below the optical diffraction limit\cite{Alden2013,Butz2014,Ju2015,Kim2015,Yin2016}. 

\begin{figure*}
    \centering
    \includegraphics[width = 180mm]{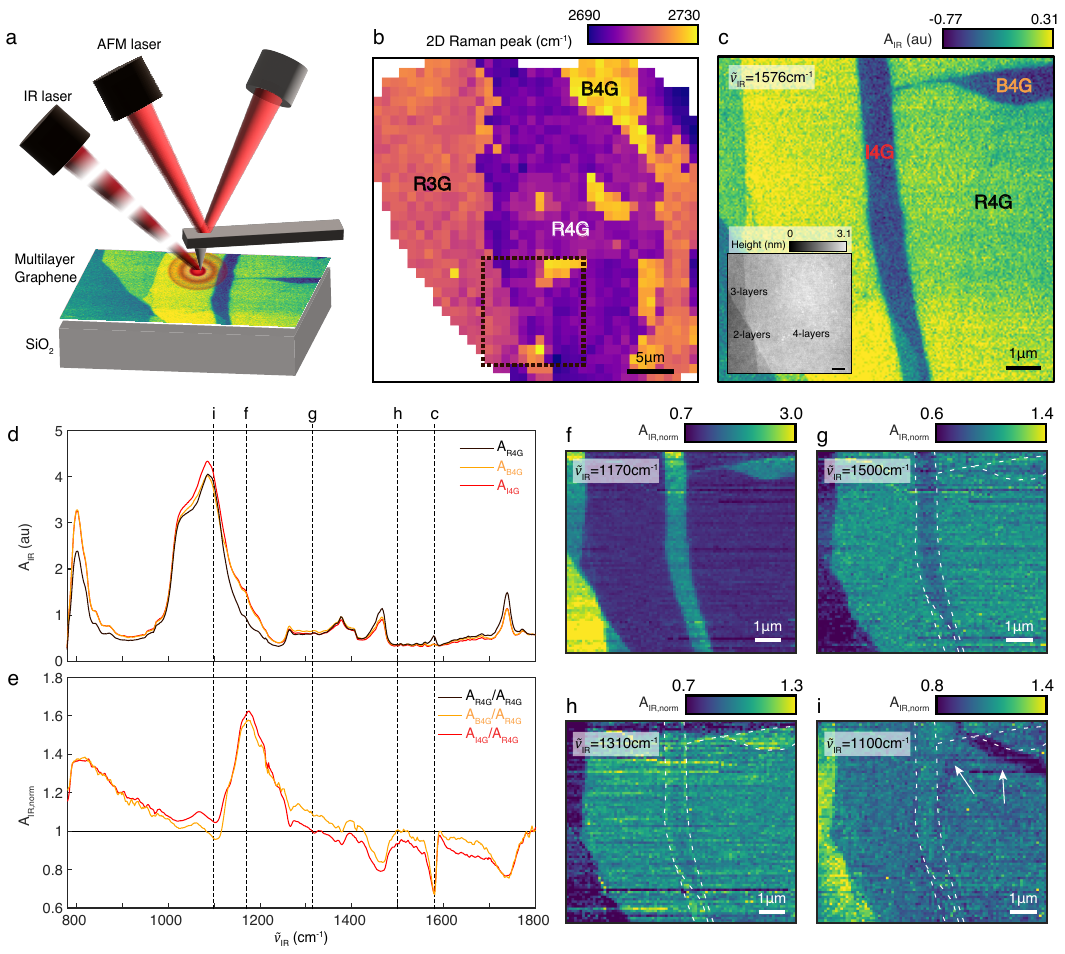}
    \caption{\textbf{AFM-IR imaging of few-layer graphene.} 
    \textbf{(a)} 
    Schematic of photothermal AFM-IR microscopy. 
    A pulsed IR laser is focused on the sample close to the tip and the resonant thermal expansion is detected by the AFM tip.
    \textbf{(b)} 
    Raman spectroscopy of a graphene flake containing two, three, and four layer region, performed using a 488nm laser. The wavenumber at the maximum of the 2D Raman peak is plotted to identify the stacking orders. 
    \textbf{(c)} AFM-IR scan of the region outlined by the dashed box in panel b. The inset shows the same area in topographic contrast.
    \textbf{(d)} IR wavenumber $\tilde{\nu}_{IR}$ dependence of the AFM-IR amplitude for the three stacking orders in four-layer graphene.
    \textbf{(e)} Same spectra normalized by the R4G spectral data. Further analysis details available in Fig.\ref{figS:spectral_analysis}.
    \textbf{(f)}AFM-IR scan at $\tilde{\nu}_{IR}$ = 1170 $cm^{-1}$
    \textbf{(g)} 1500 $cm^{-1}$
    \textbf{(h)} 1310 $cm^{-1}$ and
    \textbf{(i)} 1100 $cm^{-1}$, extracted from the  hyperspectral map (see additional data in Fig.~\ref{figS:hyperspectral_scans}).
    The dashed lines are a guide to the eye indicating  different stacking order transitions; white arrows indicate features likely associated with strain.}
    \label{fig:1}
\end{figure*}

The desirable properties for new AFM based techniques include (1) high spatial resolution, (2) high throughput, (3) unambiguous stacking order identification and (4) ability to image both exposed and hBN encapsulated graphite flakes. 
Prior work has established atomic force microscopy (AFM) based techniques including  scattering scanning near field optical microscopy (s-SNOM)\cite{Ju2015,Kim2015,Jiang2018,Wirth2022,Beitner2023,Bruker2023} and Kelvin Probe Force Microscopy\cite{Yu2021} capable of stacking order characterization, but these are suboptimal on one or more of these dimensions. 
Here we demonstrate two techniques to identify stacking orders in few layer graphene flakes and apply it to characterize the microscopic structure of tetralayer graphene.
The first technique, photothermal AFM-IR microscopy\cite{Dazzi2005}, provides absolute contrast between stacking orders and can distinguish between rhombohedral, Bernal, and intermediate stacking orders via spectral analysis.
The second technique, scanning microwave impedance microscopy (sMIM)\cite{Anlage2007,Lai2008,Lai2008a,Lai2009,Lai2010,ZhuangqunHuang2016}, shows higher contrast, and we show that it readily resolves stacking faults with sub-20nm resolution.
Both AFM-IR as well as sMIM can be used for subsurface imaging when the graphene flake is buried underneath a hexagonal-Boron Nitride (hBN) dielectric, allowing for characterization of assembled van der Waals heterostructures.

\section{Photothermal infrared microscopy}

In photothermal AFM-IR, the sample is irradiated by a pulsed infra-red laser focused in the vicinity of the AFM tip (here, a PR-UM-TnIR-D tip with a radius of 20nm).
AFM-IR detects the local thermal expansion of the sample due to the resonant absorption of the infra-red radiation at the tip location (see Fig. \ref{fig:1}a).
The resonant thermal expansion exerts a transient force on the AFM tip, making AFM-IR sensitive to absorption on the scale of the tip apex\cite{Bruker2023}.
Notably, AFM-IR is distinct from s-SNOM, in which the scattered infrared light is detected; rather,  AFM-IR can be thought of a local equivalent of far field IR absorption measurements.
To test the sensitivity and resolution of AFM-IR for detecting different stacking orders, we exfoliate few layer graphene flakes onto 285nm thick SiO$_2$/Si substrates pretreated with O$_2$-plasma. 
Layer numbers were identified via optical microscopy and standard AFM topography. 
Fig. \ref{fig:1}b shows a Raman image using a 488nm laser; we characterize the domain structure by the shift in the 2D peak appearing at wavenumber near 2700 cm$^{-1}$\cite{Lui2011a,Cong2011}. We identify both rhombohedral (red to purple on the color scale) and Bernal (yellow) stacking orders.  Note that this flake has multiple wrinkles indicating the presence of considerable strain (see Fig. \ref{figS:largeArea}): this likely contributes to the rich stacking domain structure we observe.  
Fig.~\ref{fig:1}c shows AFM-IR microscopy at a fixed excitation $\tilde{\nu}_{IR}$ = 1576 cm$^{-1}$, taken in a smaller region consisting mostly of tetralayer graphene (black dashed box in Fig. \ref{fig:1}b). We observe large AFM-IR contrast within the atomically smooth region, showing  features corresponding to the Bernal area and large areas of rhombohedral stacking and a region with a different contrast cutting vertically through the center of the image (see also Fig. \ref{figS:largeArea}).
We identify  Bernal (B4G, e.g. ABAB) and rhombohedral (R4G, e.g. ABCA) areas from the Raman scan; in addition, a third dark region--manifesting as a central strip in Fig. \ref{fig:1}c---shows the same contrast as B4G in AFM-IR but the same contrast as R4G in Raman,  We assign this domain to the intermediate stacking order (I4G, e.g. ABCB), the third possible metastable crystal structure possible in four layer graphene\cite{Wirth2022,Beitner2023}. 

We corroborate these assignments via spectral analysis. 
AFM-IR spectra obtained by keeping the tip position fixed and sweeping the wavelength of the infrared laser are shown in Fig. \ref{fig:1}d in the  R4G, B4G and I4G stacking regions, respectively.
The most prominent features of the spectrum are two large peaks around $\tilde{\nu}_{IR}$ = 800 cm$^{-1}$ and 1100 cm$^{-1}$, which we attribute to Si-O bonds likely arising from the underlying SiO$_2$ substrate\cite{Bruker2023}.  Additional peaks at higher frequency arise from the graphene layers, and are strongly dependent on stacking order. 
In particular, rhombohedral stacking order can be readily identified by the presence of a clear peak at $\sim$1580cm$^{-1}$.  
To find more IR-wave numbers that are sensitive to stacking order, we acquire a hyperspectral image of 100x100 points over a 10x10$\mu m$ area by sweeping $\tilde{\nu}_{IR}$ from 800 to 1800 cm$^{-1}$ at each point. 
We subtract a background line by line to get rid of amplitude drifts during the long measurement and normalize the amplitude by the signal measured in a confirmed R4G region (for details on the analysis, see supplementary information and Fig.\ref{figS:spectral_analysis}).
The normalized AFM-IR amplitude, A$_{IR,norm}$, is shown in Fig. \ref{fig:1}e as a function of wavenumber.
Almost all wavenumbers show contrast between stacking orders. This is somewhat surprising, as the corresponding energies of 0.1 to 0.22 eV are below the energy separation between the lowest two conduction or valence bands\cite{Mak2010}.  
This suggests that details in the lowest energy bands themselves or differences in the coefficients in thermal expansion most likely account for the differences across the AFM-IR spectra.

The largest relative contrast  between stacking orders occurs on the  shoulder of the Si-O peak,  $\tilde{\nu}_{IR} \approx$ 1170 cm$^{-1}$, corresponding to an energy of 0.145 eV and $\tilde{\nu}_{IR}$ = 1576 cm$^{-1}$ (energy of 0.195eV).
AFM-IR images at these wavenumbers are shown in Figs. \ref{fig:1}c and f, respectively.
These wavenumbers are excellent for identifying rhombohedral stacking orders, however, the contrast between B4G and I4G is minimal.
In comparison, for $\tilde{\nu}_{IR}$ = 1500 cm$^{-1}$, the I4G order shows high contrast compared to the other two stacking orders (Fig. \ref{fig:1}g), while at $\tilde{\nu}_{IR}$ = 1310 cm$^{-1}$ B4G stacking is distinct from the other orders (Fig. \ref{fig:1}h).
From this analysis, we can unambiguously assign the three different stacking orders to the different visible regions. 
The hyperspectral map contains many more wavenumbers which can help distinguish between stacking orders, and images over the whole range of wavenumbers used can be found in Fig. \ref{figS:hyperspectral_scans}.
Notably, around $\tilde{\nu}_{IR}$ = 1100 cm$^{-1}$ an additional feature seemingly uncorrelated to stacking order appears.
In Fig. \ref{fig:1}i, two dark features (indicated by white arrows) are visible near a boundary between B4G and R4G regions.
We speculate that these may be related to strain within the graphene layers; in particular, 
compressive uniaxial as well as shear strain has been shown to favor rhombohedral stacking orders in graphene\cite{Yang2019,Nery2020,Dey2024}. 
Rhombohedral and intermediate stacking orders could therefore be stabilized against relaxation to the thermodynamically more stable Bernal counterpart by pinning, leading to strain within the flake. 
\\

\begin{figure*}
    \centering
    \includegraphics[width = 180mm]{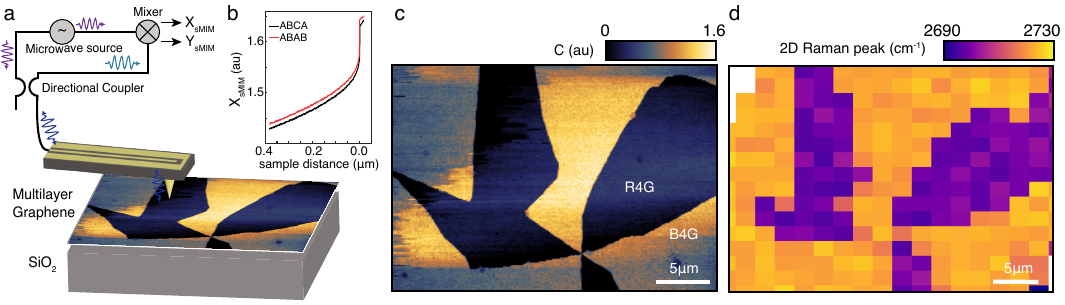}
    \caption{\textbf{sMIM imaging of few layer graphehne:} 
    \textbf{(a)} Schematic illustration of the sMIM measurement, which detects the transmission line integrated into the cantilever and terminated by a coaxial waveguide at the apex of the probe tip.  X$_{sMIM}$ and Y$_{sMIM}$ denote the two quadratures of the reflected signal. 
    \textbf{(b)} X$_{sMIM}$ signal versus tip-to-sample distance as the tip approaches the surface; the phase of the reference signal is adjusted such that Y$_{sMIM}\approx 0$. 
    \textbf{(c)} sMIM scan of an atomically smooth four-layer graphene flake. 
    \textbf{(d)} Raman spectroscopy of the same area from which we identify the stacking order via the maximum of the 2D Raman peak.
    }
    \label{fig:2}
\end{figure*}

\section{Microwave impedance microscopy}
High contrast imaging of stacking orders is also possible using scanning microwave impedance microscopy (sMIM).
In sMIM, a microwave signal is sent via coaxial cables and a transmission line within the AFM cantilever to the apex of the tip\cite{Lai2008,Lai2008a,Lai2009}. 
The reflected microwave signal is then detected as schematically depicted in Fig.\ref{fig:2}a, 
generating contrast that depends on the tip-sample impedance. 
When the tip-sample separation is smaller than the  tip diameter, contrast is dominated by the electrical properties of the sample---such as conductivity and permittivity---in the immediate region of the tip\cite{Lai2008,Lai2008a,Lai2010,ZhuangqunHuang2016}.
The reflected signal is demodulated with a mixer, giving the in- and out of phase quadratures (X$_{sMIM}$ and Y$_{sMIM}$).
The phase of the input signal is adjusted to null out Y$_{sMIM}$. 
To remove the ambiguity in the sign of X$_{sMIM}$, we ramp the tip towards and away from the surface, monitoring X$_{sMIM}$ as shown in Fig.~\ref{fig:2}a. We choose the phase of the input signal so that X$_{sMIM}$ increases as the surface is approached (Fig. \ref{fig:2}b) while Y$_{sMIM}$ remains zero.  The measured X$_{sMIM}$ then gives a consistent sign of the contrast between different stacking orders, independent of details of the tip-sample impedance arising from changes in tip shape or tip-surface interaction.
A sMIM scan of a different tetralayer graphene flake in this configuration is shown in Fig.~\ref{fig:2}c.  
As the sMIM contrast is proportional to the admittance for insulating samples, in the following we refer to it as $C$ in line with previous literature\cite{ZhuangqunHuang2016}.
We observe distinct regions with two levels of contrast, which Raman scans of the same area (Fig.~\ref{fig:2}d) confirm to be associated with R4G and B4G stacking orders. 
Combining the ramp curves taken in the two different regions and spatial scan, we can assign the lower and higher sMIM amplitude $C$ to rhombohedral and Bernal stacking orders, respectively. 
We note that sMIM has the disadvantage relative to AFM-IR of not providing an absolute measurement of the stacking order, so that the stacking order in a uniform sample of a single stacking order cannot be determined absent an additional calibration such as Raman spectroscopy. However, the considerably larger signal to noise and lower sensitivity to tip or surface preparation and contamination make it an ideal tool for characterizing mixed domain samples. 

\section{Subsurface imaging of stacking orders}
\begin{figure}[t]
    \centering
    \includegraphics[width=\columnwidth]{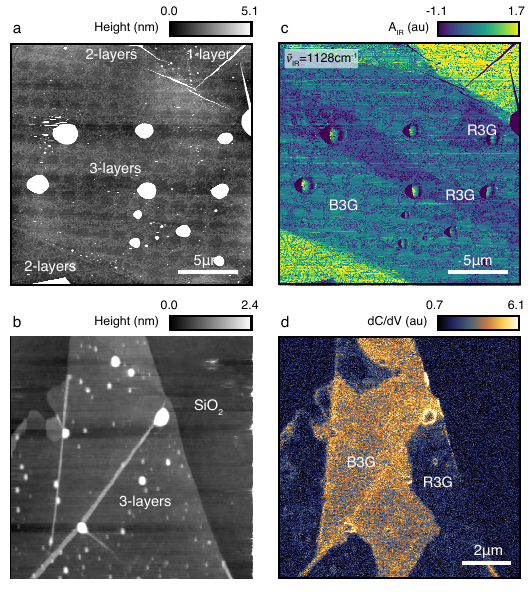}
    \caption{\textbf{Subsurface imaging of graphene stacking orders.}
    \textbf{(a)}-\textbf{(b)} AFM topography scans of two hBN-encapsulated trilayer graphene samples.
    \textbf{(c)} AFM-IR scan at $\tilde{\nu}_{IR}$ = 1128cm$^{-1}$. 
    The stacking orders are labeled. No corresponding feature is seen in topography. 
    \textbf{(d)} sMIM dC/dV scan of the heterostructure in panel b with a 5V AC amplitude applied to the tip.
    }
    \label{fig:3}
\end{figure}

Both AFM-IR and sMIM imaging do not rely on direct electrical contact and so are well suited to imaging layers already incorporated into van der Waals heterostructures. To demonstrate this, we study trilayer flakes containing both rhombohedral and Bernal stacked regions, covered in 5-10nm of hexagonal boron nitride dielectric.  Devices are fabricated using standard polycarbonate-based dry transfer techniques, with the polymer dissolved in chloroform following delamination\cite{Wang2013}.
Fig.\ref{fig:3}a-b shows topographic scans of the resulting heterostructures; the trilayer is visible through the hBN, along with bubbles formed by contaminants trapped between the hBN dielectric and the graphene layers as well as polymer residue at the surface. 
Fig. \ref{fig:3}c shows an AFM-IR scan at $\tilde{\nu}_{IR}$ = 1128cm$^{-1}$ in which we observe both rhombohedral and Bernal stacking orders.  
Notably, the IR contrast is similar to unencapsulated samples.
For example at $\tilde{\nu}_{IR}$ = 1128cm$^{-1}$, R3G is darker than B3G, with the contrast reversed at $\tilde{\nu}_{IR}$ = 1576 cm$^{-1}$ (see Fig.\ref{figS:IRsubsurface}).
Apparently,  the  thermal expansion of the graphene layer due to resonant IR heating is unaffected by the intermediate hBN cladding layer.
As the hBN band-gap is much larger than the IR-photon energies, there is no direct heating of the hBN by the infrared laser, and we expect similar contrast for graphene buried under thicker hBN layers, albeit at the cost of diminished spatial resolution. 

\begin{figure*}
    \centering
    \includegraphics[width = 180mm]{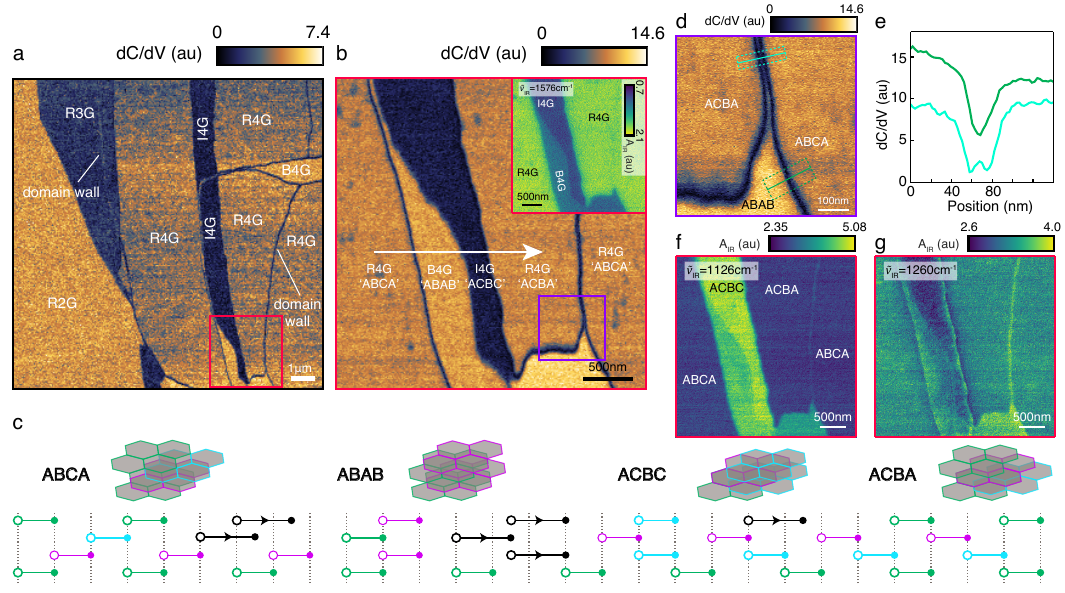}
    \caption{\textbf{Stacking order transitions and domain walls.}
    \textbf{(a)} $dC/dV$ sMIM scan of a similar area as shown in Fig.~\ref{fig:1}c.
    \textbf{(b)} Detailed view of region marked by the red rectangle in panel a. 
    Inset: same region imaged with AFM-IR at $\tilde{\nu}_{IR}$ = 1576cm$^{-1}$.
    \textbf{(c)} Schematic illustration of  stacking order transitions separating  ABCA (rhombohedral) and ACBA (intermediate) regions along the white arrow marked in panel b. 
    The stretched unit cells (black) indicate domain walls developing in response to shear strain between top and bottom of the flake; they shift the top one, two or three layers.
    Here green, pink and cyan indicate layers at the `A', `B' or `C' position within the multilayer graphene lattice, respectively.
    \textbf{(d)} High magnification scan using sMIM in the area outlined in  purple box in panel b.
    \textbf{(e)} 
    $dC/dV$ amplitude across the single and double domain wall feature marked in green and cyan in panel d, respectively. The traces are averaged over the area indicated for better signal to noise.
    \textbf{(f)} AFM-IR scans of the same region as panel b at $\tilde{\nu}_{IR}$ = 1126cm$^{-1}$ and \textbf{(g)} $\tilde{\nu}_{IR}$ = 1260cm$^{-1}$.
    The contrast at these wavenumbers is not consistent with only three stacking orders, suggesting a role for uniformly distributed strain.  } 
    \label{fig:4}
\end{figure*}

Similarly, sMIM imaging shows significant contrast in hBN clad layers, with two different contrast regions visible in  (Fig.~\ref{fig:3}d) which we attribute to ABA and ABC stacking orders.  
We note that this measurement is performed in $dC/dV$ mode, in which the voltage on the metallic sMIM tip is varied at a finite frequency (typically near 80kHz) and the demodulated radio frequency signal is read-out via lock-in techniques at the modulation frequency.   
Notably, we did not observe any contrast in the unmodulated signal measured at the same time while for unencapsulated samples, both $C$ and $dC/dV$ show good contrast ((see Figs.~\ref{figS:sMIMburied} and ~\ref{figS:dCdV}). 
A possible culprit might be the small AC electric field produced by the microwave signal when the tip is only weakly capacitively coupled through the hBN to the graphene layer. 
Additional modulation significantly enhances the contrast, as previously shown for other semiconductors\cite{ZhuangqunHuang2016}.

In the future, it might be worth extending these techniques to samples encapsulated by both hBN dielectrics and metallic graphite gates, the typical geometry used in dual-gated devices.
For this purpose, AFM-IR would likely be the preferable choice, as its typical depth sensitivities of $\sim$50nm for IR-active materials\cite{Bruker2023} are within the range of the thickness of typical van der Waals heterostructures. 
This would lead to the resonant absorption and thermal expansion that can be detected with AFM-IR.
In contrast, sMIM contrast suffers significantly from the effects of metallic screening layers.

\section{Imaging of domain wall defects}

Graphene multilayers may also feature domain walls between regions of \textit{identical} stacking order.  
These features are associated with shifts in the registry between two adjacent layers, with the stacking order returning to the the original one over the course of 5-10 nm\cite{Alden2013,Butz2014}, and were previously studied most extensively in Bernal bilayer graphene \cite{Ju2015,Yin2016,Jiang2016}. Practically, domain walls may interfere with device fabrication when isolating uniform stacking order is desirable. As high resolution scanning probes, both AFM-IR and sMIM are well positioned to solve this problem via high throughput nano-scale imaging. Indeed, as shown in Fig.~\ref{figS:BBGF_domainwalls}, sMIM is able to visualize bilayer graphene domain walls with high contrast. 

Fig. \ref{fig:4}a shows a sMIM image of the same region characterized by AFM-IR in Fig. \ref{fig:1}.  Domain walls within otherwise uniform regions of stacking order are readily visible, manifesting as dark lines. 
We further observe a difference in contrast for three different stacking orders, identified based on our AFM-IR analysis as rhombohedral, Bernal, and intermediate stacking. 
Interestingly, we find different regions of nominally identical rhombohedral stacking order to be separated by both  domain walls as well as macroscopic regions of intermediate or Bernal stacking.  
To understand the evolution of the stacking order in space, we analyze a smaller, $3.2\mu m\times 3.2 \mu m$ subregion indicated by the red rectangle in Fig. \ref{fig:4}a.  
The corresponding sMIM image is shown in Fig.~\ref{fig:4}b.  Our assignments of R4G, B4G, and I4G stacking is confirmed by AFM-IR measurements at $\tilde{\nu}_{IR}$ = 1576cm$^{-1}$, shown in the inset.  

Assignments of stacking order based on bulk spectroscopy are, of course, not unique---domains related by translation symmetry or mirror symmetry about the center plane of the graphite, are expected to show identical contrast.  However, analysis of the domain wall pattern provides significant information about the relative shifts between layers and the corresponding underlying strain patterns in the sample.  An example of a consistent assignment of the individual domain wall structure is given in schematically in Fig. \ref{fig:4}c for the trajectory indicated by the white arrow in Fig. \ref{fig:4}b.  
For simplicity, we keep the bottom layer fixed at the `A' position (green) and allow for shifts of the top one, two or three layers between `A', `B' (pink) and `C' (cyan) positions in the same direction; we define the left most region as `ABCA' stacking. 

The shift to ABAB B4G stacking can be understood as a shift by a single nearest-neighbor distance of the top two layers of ABCA relative to the bottom two. 
The next transition to ACBC I4G stacking corresponds to a shift of the top three layers in the same direction, and the final transition to ACBA stacking corresponds to a shift of the top layer by 1/3 unit cell, again in the same direction. 
This pattern of shifts is consistent with a shear strain pattern arising from bending of the graphite flake, with 1, 2, or 3 fewer atoms in the 2nd, 3rd, or 4th layers relative to the bottom layer across the several micron extent of the marked trajectory.   

The presence of macroscopic regions of I4G and R4G stacking is indicative of significant pinning, as these stacking orders are higher energy than the B4G stacking and so could be expected to contract to the atomic scale if the lattice is allowed to relax.   
Notably, however, when such domains correspond to a net shift in number of atoms in a given layer, they cannot contract completely.  Evidence for this is shown in Fig. \ref{fig:4}d-e, which shows a detail of a region in which two R4G regions are separated by a narrow domain wall. 
A high resolution scan across this domain wall is shown Fig. \ref{fig:4}e, revealing internal structure at the 10nm scale. 
Indeed, in the bottom of Fig. \ref{fig:4}d, this domain wall expands into a macroscopic B4G region.
Tracing the left branch of the domain wall in Fig. \ref{fig:4}b, moreover, we associated it with a contracted region of I4G stacking.  
We can thus conclude that the double domain wall rendered in dark blue in Figs. \ref{fig:4}d corresponds to the ABAB-ACBC transition region between the ACBA and ABCA domains, contracted to the few nm scale.  

While our assignment of stacking order transitions is consistent across the domain boundaries discussed, the picture can get be more complicated when considering strain over micrometer scale.
We find evidence for such in AFM-IR scans at wavenumbers that are more sensitive to strain, as shown in Fig.\ref{fig:4}f,g.
Additional contrast and gradients in the IR-amplitude appear inconsistent with the simple picture of the three possible stacking orders in tetralayer graphene, likely an indication of strain on macroscopic length scales, in which complete relaxation to a metastable stacking order has not occurred.

In summary, we employed AFM-IR and sMIM to characterize graphite allotropes and demonstrated their suitability for nanoscale imaging and high yield and throughput sample fabrication of single domain multilayer graphene devices.
Both techniques offer several advantages compared to known and other possible AFM-based imaging techniques - such as subsurface imaging, reliable high signal to noise and good spatial resolution (see supplement and Fig.~\ref{figS:topo},\ref{figS:conductive} for additional testing of other AFM techniques). Our results demonstrate that AFM-IR and sMIM are ideally suited both improving fabrication yields in metastable stacking order as well as investigations of dynamical stacking domain rearrangements such as those that occur under applied electric fields\cite{Li2020,Singh2025}, strain, or heating.

\textbf{Acknowledgements.}
This work was supported primarily by the Air Force 
Office of Scientific Research under award \#FA9550-24-1-0113 to AFY.  
OIS acknowledges direct support by the National Science Foundation through Enabling Quantum Leap: Convergent Accelerated Discovery Foundries for Quantum Materials Science, Engineering and Information (Q-AMASE-i) award number DMR-1906325; the work also made use of shared equipment sponsored by under this award. 
Additional support was provided by a Brown Investigator Award to AFY. 
KW and TT acknowledge support from the JSPS KAKENHI (Grant Numbers 21H05233 and 23H02052) and World Premier International Research Center Initiative (WPI), MEXT, Japan.

\textbf{Author Contributions} P.D.W. and A.F.Y. conceived and directed the project. L.H., C.L.P, O.S., Youngjoon Choi, Q.H., J.K. and C.P. performed the AFM-IR measurements. L.H., L.C., N.S., M.V. and P.D.W. performed the sMIM measurements. L.H., Ysun Choi, Youngjoon Choi, J.Z. and H.S. prepared the samples. L.H. and A.F.Y. wrote the manuscript with input from all other authors. 

\bibliographystyle{apsrev4-1}
\bibliography{bibliography}


\clearpage
\newpage
\pagebreak

\onecolumngrid


\setcounter{equation}{0}
\setcounter{figure}{0}
\setcounter{table}{0}
\setcounter{section}{0}
\makeatletter
\renewcommand{\theequation}{S\arabic{equation}}
\renewcommand{\thefigure}{S\arabic{figure}}
\renewcommand{\thepage}{\arabic{page}}

\textbf{Supplementary information for 'Nanoscale infrared and microwave imaging of stacking faults in multilayer graphene':}\\
This supplementary file includes:
\begin{itemize}
    \item Sample fabrication details 
    \item Method details on the different AFM techniques used
    \item Supplementary Figures
\end{itemize}

\textbf{Sample preparation:}
Multilayer graphene flakes were exfoliated onto silicon wafers with 285nm thickness of SiO$_2$. Notably, we obtained good AFM-IR contrast only for multilayer graphene flakes exfoliated onto SiO$_2$ substrates that were pre-treated with oxygen plasma immediately beforehand.
The graphene structures were prepared by dry transfer of a hexagonal boron nitride (hBN) dielectric layer\cite{Wang2013} using polycarbonate membranes onto the trilayer graphene flake, followed by removal of the polymer with chloroform. The periodic pattern of circular features observed in both topography and AFM-IR in \ref{fig:3} originates from a prior Raman scan, which burned residual polymer on the sample surface.

\begin{figure*}
    \centering
    \includegraphics[width = 180mm]{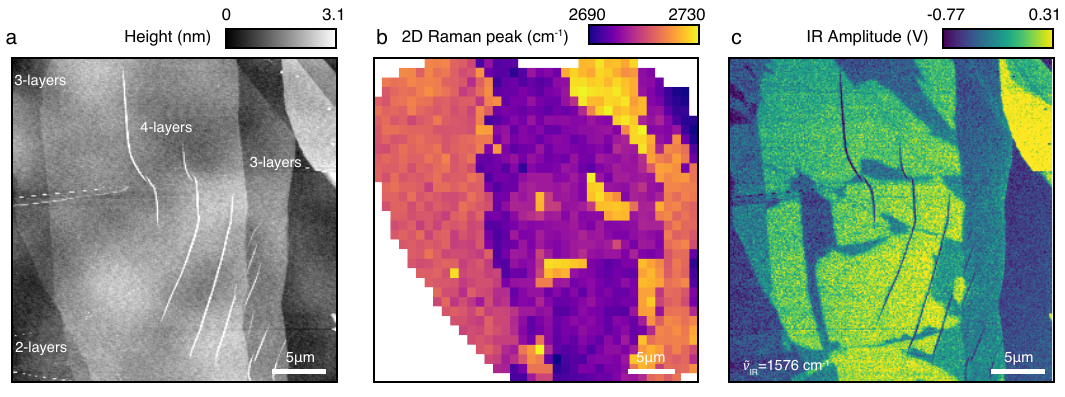}
    \caption{\textbf{Large area view of AFM-IR images for a full comparison to topography and Raman:}
    \textbf{(a)} topographical scan which shows significant wrinkles within the sample which indicates strain might be trapped in the sample.
    \textbf{(b)} Raman map of roughly the same area (same as main text Fig.~\ref{fig:1}b).
     \textbf{(c)} AFM-IR scan of the same area as in panel a. Apart from contrast also observed in topography, features corresponding to stacking order changes appear consistent with the Raman map.
    }
    \label{figS:largeArea}
\end{figure*}

\textbf{Raman microscopy:}
Raman maps are performed with a 488nm laser focused on a spot the size of $\sim$ 1 $\mu$m. The color maps in Fig.\ref{fig:1},\ref{fig:2},\ref{figS:largeArea} were generated by extracting the maximum of the 2D Raman peak.

\textbf{Photothermal AFM-IR:}
AFM-IR measurements were performed using a commercially available Dimension IconIR system from Bruker Nano, which offers a topographic noise floor of 30 picometers (rms) and a thermal drift of 0.2 nm/min. The system was coupled with a multi-chip quantum cascade laser (QCL) source (MIRcat, Daylight Solutions), with a typical 1mW IR laser power focused beneath the AFM cantilever. The laser covers the mid-IR range from 1800 to 800 cm$^{-1}$, with a spectral resolution of 0.1 cm$^{-1}$. The data were acquired in tapping AFM-IR mode using a PR-UM-TnIR-D probe with a 42 N/m spring constant and a 20 nm tip radius. The probe was driven at its second resonance frequency (1.5 MHz), while detection occurred at the first resonance (250 kHz), with the IR pulse rate tuned to their frequency difference (1.25 MHz). Typical AFM-IR spectrum was acquired in 10 seconds at a laser sweep rate of 100 cm$^{-1}$/s. A hyperspectral image of 100x100 points over a 10x10 $\mu$m area was acquired overnight by recording a full range spectrum at each point in 5 seconds.

\textbf{Scanning Microwave Impedance Microscopy:}
The sMIM measurements are performed on a commercially available Dimension Icon from Bruker Nano equipped with the ICONSMIM-PRO add-on. Shielded sMIM probes (SMIM-150 from Bruker) were aligned using a reflectometer to achieve a resonance with attenuation of -20dB to <-30dB with respect to the output signal. 
During the approach monitoring $X_{sMIM}$ and $Y_{sMIM}$, the phase is adjusted such that $Y_{sMIM}$=0 and $X_{sMIM}$ positive. This ensures a consistent relative amplitude contrast between rhombohedral, Bernal and intermediate stacking domains.

\textbf{Topographical AFM imaging of graphene stacking orders:} 
Interestingly, we find that tapping-mode atomic force microscopy can be sensitive to graphene stacking orders under certain conditions. For instance, in Fig.~\ref{figS:topo}, we image a pentalayer graphene flake exhibiting both Bernal and rhombohedral stacking, as identified by infrared imaging\cite{Lu2024,Feng2024} (panel b).

We use a high spring constant tip (e.g., Arrow-NCPt-50, NanoWorld) with a high drive amplitude on a Bruker Dimension Icon system. Features in topography (Fig.~\ref{figS:topo}) correspond to the stacking orders identified in the IR image. For pentalayer graphene, we observe a height difference of approximately 0.5nm between the stacking domains along the red line in panel c. This contrast may arise from differences in the elastic moduli of the respective crystal structures, which affect the force on the AFM tip in tapping made. However, this contrast mechanism is not stable over several scans and therefore not suitable for reliable identification of stacking orders.

\textbf{Conductive AFM imaging of graphene stacking orders:}
Conductive AFM has been used to image moiré structures in twisted graphene sheets\cite{Zhang2020, Huang2021, Kim2023}. In Fig.~\ref{figS:conductive}, we apply this technique to a trilayer graphene flake contacted by a dropped gold lead, as described in ref. \cite{Kim2023}, to avoid surface contamination.

Measurements were taken using a Cypher AFM system (Asylum Research) with the Orca module and an Arrow-NCPt-50 tip (NanoWorld) in an argon glovebox. Several domains and domain walls appear with good contrast (Fig.~\ref{figS:conductive}) in an otherwise atomically smooth region.

However, we note key limitations of conductive AFM compared to sMIM. Conductive AFM requires electrical contact to the sample, necessitating an exposed graphene surface and additional fabrication steps that must avoid contamination. Reliable tip-sample contact is crucial but can be easily disrupted by topographical features (e.g., the sample edge in Fig.~\ref{figS:conductive}b) or contaminants. Furthermore, current contrast varies from scan to scan and cannot be reliably assigned to specific stacking orders, requiring complementary tools such as Raman spectroscopy for each measurement. These issues make conductive AFM impractical for high-throughput sample characterization and highlight the advantages of sMIM in studying the electrical properties of multilayer graphene.

\begin{figure*}
    \centering
    \includegraphics[width = 120mm]{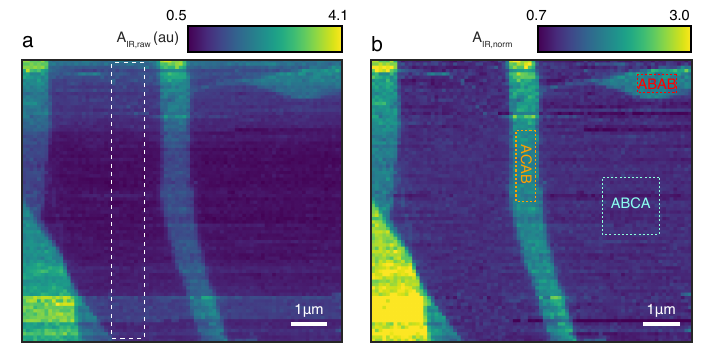}
    \caption{\textbf{Data analysis of the hyperspectral map:} 
    \textbf{(a)} single raw image slice of the hyperspectral data at $\tilde{\nu}_{IR}$ = 1170 cm$^{-1}$. Due to the length of the measurement (approximately 16 hours), drifts in the raw signal, $A_{IR,raw}$ appear. To remove the drifts, we subtract a background line by line by taking the average value within the white dashed box which consist of pure ABCA stacking order throughout the vertical axis of the image. We refer to the result as $A_{IR,woBG}$.
    To quantify the contrast at different wavenumbers, we normalize the data.
    We add the average value of an ABCA region to $A_{IR,woBG}$ and divide by the same value. This will normalize the signal $A_{IR,norm}$ to unity in ABCA stacked regions with contrast to the other stacking orders.
    \textbf{(b)} normalized signal $A_{IR,norm}$ of the same hyperspectral slice after processing. 
    The spectral cuts of $A_{IR}$ and $A_{IR,norm}$ shown in \ref{fig:1}d, e, are area averages over the dashed boxes as labeled.}
    \label{figS:spectral_analysis}
\end{figure*}

\begin{figure*}
    \centering
    \includegraphics[width = 140mm]{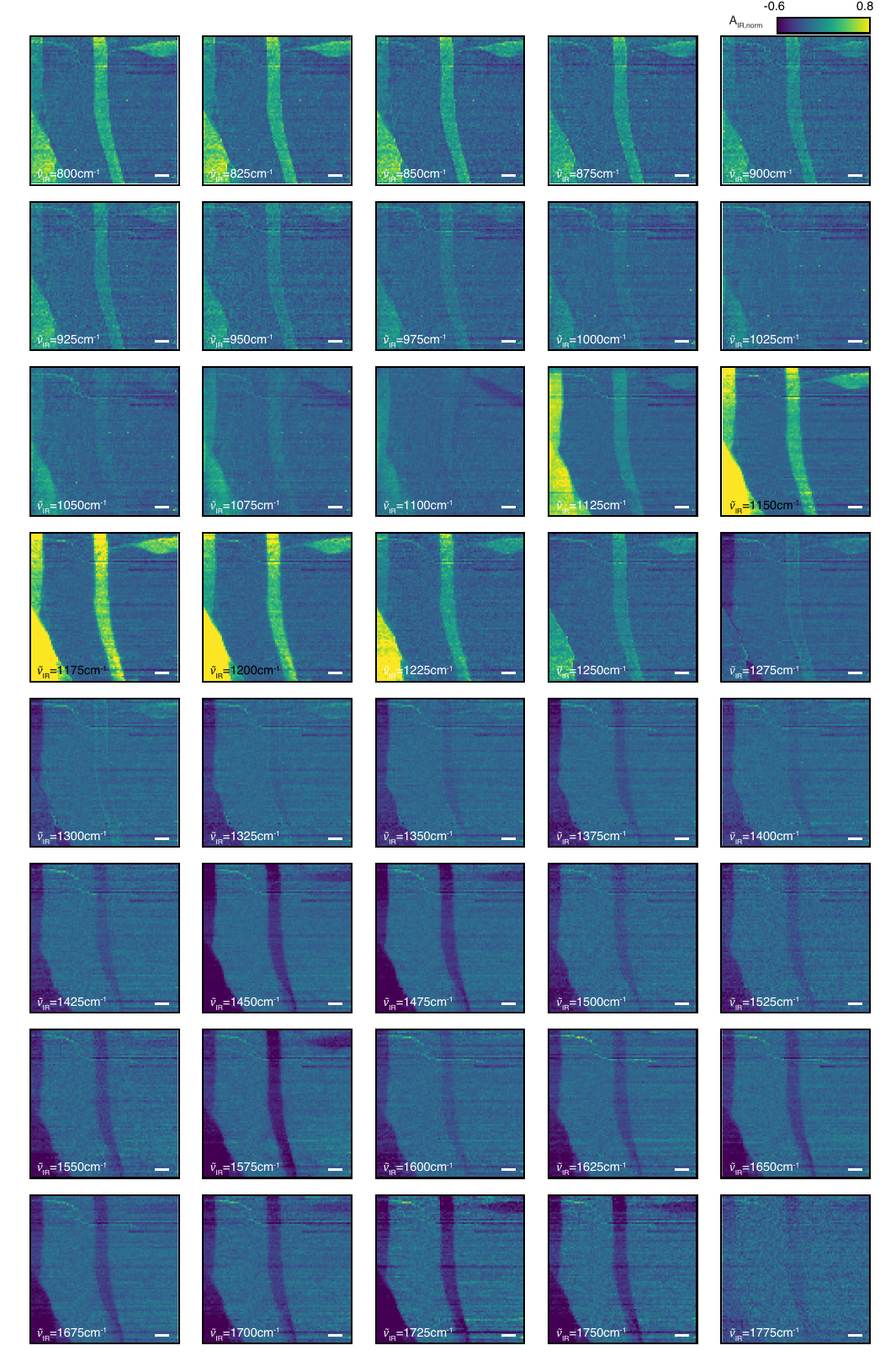}
    \caption{\textbf{Snapshots of the AFM-IR hyperspectral data for many wavenumbers $\tilde{\nu}_{IR}$:} each panel is taken at a different wavenumber as labeled. The scale bar is 1$\mu$m.} 
    \label{figS:hyperspectral_scans}
\end{figure*}

\begin{figure*}
    \centering
    \includegraphics[width = 90mm]{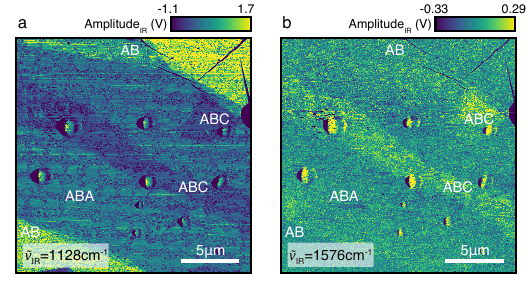}
    \caption{\textbf{AFM-IR of the hBN encapsulated sample at another wavenumber:} 
    \textbf{(a)} scan at $\tilde{\nu}_{IR}$ = 1128 cm$^{-1}$, same as in the main text.
    \textbf{(b)} scan at $\tilde{\nu}_{IR}$ = 1576 cm$^{-1}$ for comparison.
    ABA (Bernal) and ABC (rhombohedral) stacking orders are labeled.}
    \label{figS:IRsubsurface}
\end{figure*}

\begin{figure*}
    \centering
    \includegraphics[width = 180mm]{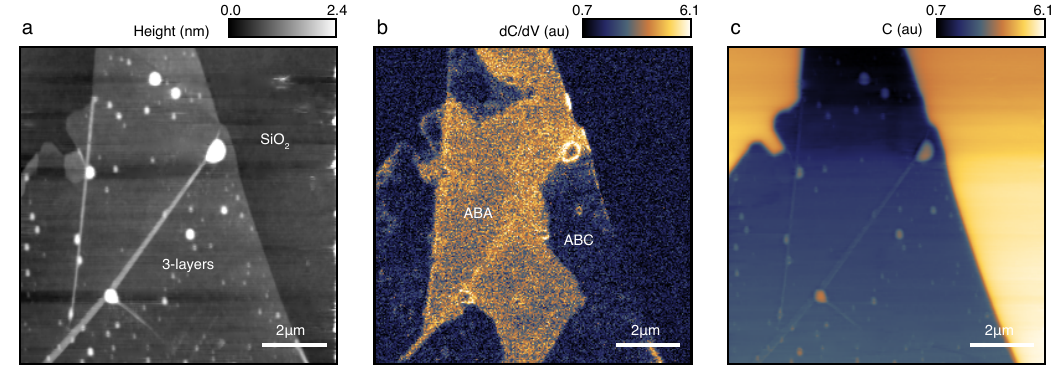}
    \caption{\textbf{Capacitance and $dC/dV$ signal of sMIM when imaging through hBN:} 
    \textbf{(a)} topographical AFM scan of a trilayer graphene encapsulated by hBN, as shown in main text Fig.~\ref{fig:3}b.
    \textbf{(b)} sMIM scan of the same area in $dC/dV$ contrast, as shown in main text Fig.~\ref{fig:3}d.
    \textbf{(c)} comparison of the sMIM signal $C$ measured at radio frequencies. No signal associated with the different stacking orders can be seen.}
    \label{figS:sMIMburied}
\end{figure*}

\begin{figure*}
    \centering
    \includegraphics[width = 180mm]{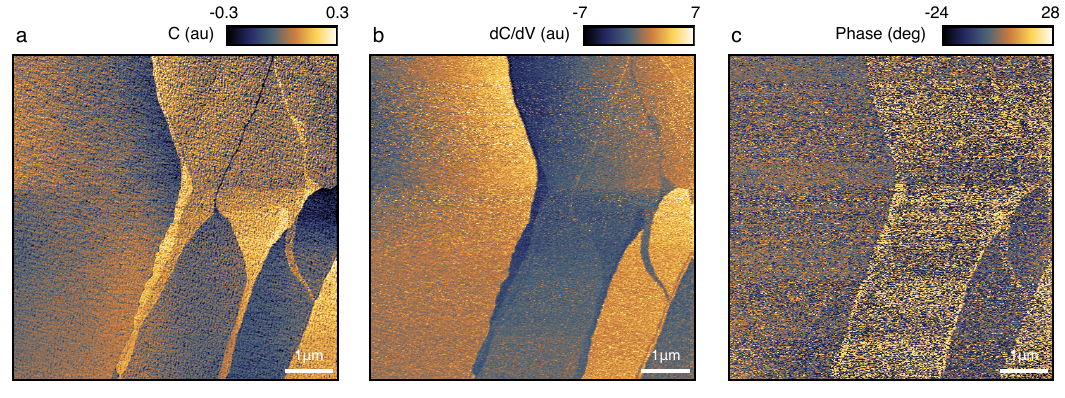}
    \caption{\textbf{Comparison of sMIM signal and AC modulated sMIM signal:} 
    \textbf{(a)} sMIM scan of a unencapsulated four-layer graphene flake when no voltage modulation is applied to the tip on top of the microwave signal, referred to as $C$.
    \textbf{(b)} sMIM scan of the same area when a 3V AC voltage is applied to the tip, referred to as $dC/dV$.
    \textbf{(c)} relative phase of the measured AC signal.}
    \label{figS:dCdV}
\end{figure*}

\begin{figure*}
    \centering
    \includegraphics[width = 180mm]{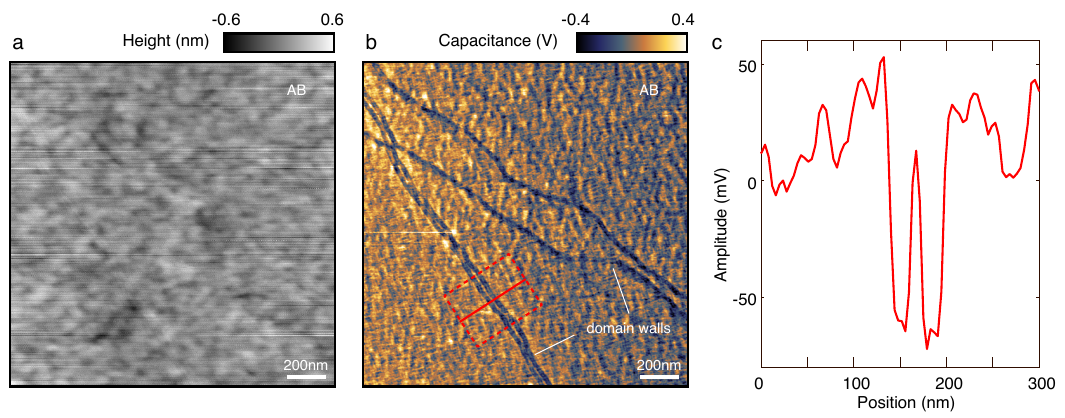}
    \caption{\textbf{sMIM images of domain walls in Bernal bilayer graphene:} 
    \textbf{(a)} topographical AFM scan of a bilayer graphene flake. The scan region is atomically flat.
    \textbf{(b)} sMIM scan of the same area. Several features run diagonally across the image which we associate with domain walls.
    \textbf{(d)} sMIM amplitude $C$ cross-sectional profile  of a double domain wall at the position marked by the red line in panel b. The dashed box indicates over which area the data is averaged to improve the signal to noise overcoming the background of the SiO$_2$ roughness.}
    \label{figS:BBGF_domainwalls}
\end{figure*}

\begin{figure*}
    \centering
    \includegraphics[width = 120mm]{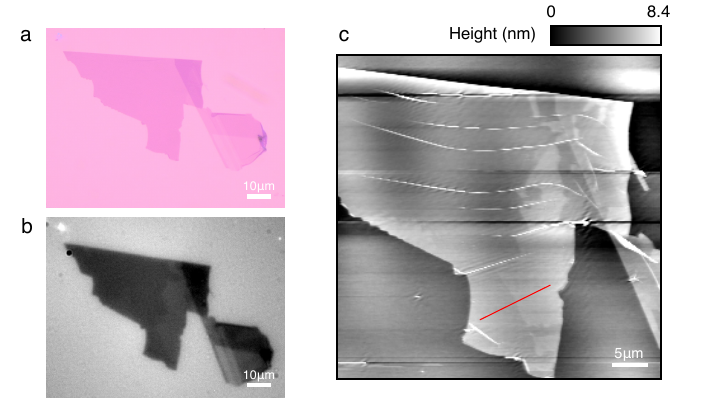}
    \caption{\textbf{Stacking orders in pentalayer graphene imaged by topographical AFM:} 
    \textbf{(a)} Optical micrograph of a pentalayer graphene flake. 
    \textbf{(b)} IR image of the same flake showing two levels of contrast which we associate with two different stacking orders \cite{Lu2024, Feng2024}.
    \textbf{(c)} Topographical AFM map. While the optical micrographs show a uniform 5-layer graphene flake, we observe a height difference between regions with different stacking orders. The topographical step along the red line is $\sim$0.5nm.}
    \label{figS:topo}
\end{figure*}

\begin{figure*}
    \centering
    \includegraphics[width = 120mm]{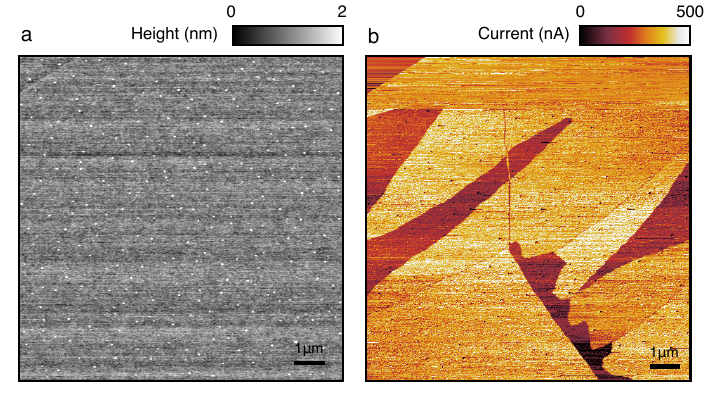}
    \caption{\textbf{Stacking orders in trilayer graphene imaged by conductive AFM:} 
    \textbf{(a)} Topographical AFM scan of a trilayer graphene flake. 
    \textbf{(b)} conductive AFM image of the same area showing different stacking orders and domain walls.}
    \label{figS:conductive}
\end{figure*}

\end{document}